\documentclass[ reprint,nofootinbib,amsmath,amssymb,aps]{revtex4-1}
\usepackage{graphicx}
\usepackage{dcolumn}
\usepackage[colorlinks,linkcolor=magenta,anchorcolor=cyan,citecolor=blue]{hyperref}
\usepackage{bm}
\usepackage[multiple]{footmisc}
\newcommand{\fig}[1]{Fig.\ref{#1}}
\def\be{\begin{equation}}
\def\ee{\end{equation}}
\def\ba{\begin{eqnarray}}
\def\ea{\end{eqnarray}}

\def\nn{\nonumber}
\def\lf{\left}
\def\rt{\right}

\newcommand{\eq}[1]{(\ref{#1})}

\def\nn{\nonumber}\def\lf{\left}\def\rt{\right} \def\w{\omega}  \def\y {\psi}   \def\p {\pi} \def\a {\alpha}  \def\d {\delta} \def\f {\phi}  \def\h {\eta} \def\j {\varphi} \def\k {\kappa} \def\l {\lambda} \def\z {\zeta} \def\x {\xi} \def\c {\chi}   \def\m {\mu} \def\pd {\partial} \def \inf {\infty}  
\def\Q{\Theta} \def\W{\Omega} \def\Y {\Psi}    \def\S {\Sigma}  \def\F {\Phi}  \def\L {\Lambda}    \def\grad{\nabla}\def\.{\cdot}
\def\math {\mathcal}
\begin{document}

\title{Action growth rate for a higher curvature gravitational theory}
\author{Jie Jiang}
\email{jiejiang@mail.bnu.edu.cn}
\affiliation{Department of Physics, Beijing Normal University, Beijing 100875, China\label{addr1}}
\date{\today}

\begin{abstract}
In this paper, we use the ``complexity equals action" (CA) conjecture to discuss the action growth rate in a black hole with multiple Killing horizons for a higher curvature theory of gravity. Based on the Noether charge formalism of Iyer and Wald, a general formalism can be resorting to finding the action growth rate within the WDW patch at the late time approximation. Moreover, as an application, we apply this formalism to a $U(1)$ invariance matter fields and utilise our results in two specific cases. Our results are universal and can be considered as the extension of the asymptotic AdS to the arbitrary asymptotic one.

\end{abstract}
\maketitle
\section{Introduction}

In recent years, there has been a growing interest in the topic of ``quantum complexity", which is defined as the minimum number of gates required to obtain a target state starting from some reference states \cite{1,2}. In the holograph viewpoint, Brown $et\,al.$ suggest that the quantum complexity of the state in the boundary theory is corresponding to some bulk gravitational quantities which are called ``holographic complexity". Then, the two conjectures, ``complexity equals volume" (CV) \cite{4,5} and ``complexity equals action" (CA) \cite{BrL,BrD}, were proposed. These conjectures have aroused researchers' widespread attention to both holograph complexity and circuit complexity in quantum field theory, $e.g.$ \cite{D,A2,A5,B1,D1,D2,D3,D4,D5,D6,D7,D8,D9,D10,D11,D12,D13,D14,D15,D16,D17,D18,D19,D20,D21,D22,D23,D24,D25,D26,F1,F2,Huang:2016fks}.

In the present work, we only focus on the CA conjecture, which states that the complexity of a particular state $|\y(t_L,t_R)\rangle$ on the AdS boundary is given by
\ba
\math{C}\lf(|\y(t_L,t_R)\rangle\rt)\equiv\frac{I_\text{WDW}}{\p\hbar}\,,
\ea
where $I_\text{WDW}$ is the on-shell action within the corresponding Wheeler-DeWitt (WDW) patch, which is enclosed by the past and future light sheets sent into the bulk spacetime from the timeslices $t_L$ and $t_R$ on the asymptotic boundary. It has been pointed out in \cite{4,5} that the late time approximation of the action growth rate should satisfy the following bound conditions
\ba\begin{aligned}
&\text{uncharged}:  \frac{d I_\text{WDW}}{d t}\leq2M\,,\\
&\text{charged}:  \frac{d I_\text{WDW}}{d t}\leq2\lf[(M-\m Q)-\lf(M-\m Q\rt)_\text{gs}\rt]\,,\\
&\text{rotating}:  \frac{d I_\text{WDW}}{d t}\leq2\lf[(M-\W J)-\lf(M-\W J\rt)_\text{gs}\rt]\,,
\end{aligned}\ea
where $\W$ and $\m$ are the angular velocity and chemical potential of the black holes, and the parameters $M,J$ and $Q$ are the black hole mass, angular momentum, and charge, respectively. The subscript ``gs" denotes the ground state of the black hole. These conditions play a crucial role in testing the compatibility and preciseness of the CA conjecture.
The first step to accomplish this test is computing the action growth rate. Hence, It is necessary for us to obtain a general formalism of the late time action growth rate in a stationary black hole.

\begin{figure*}
\centering
\includegraphics[width=0.8\textwidth]{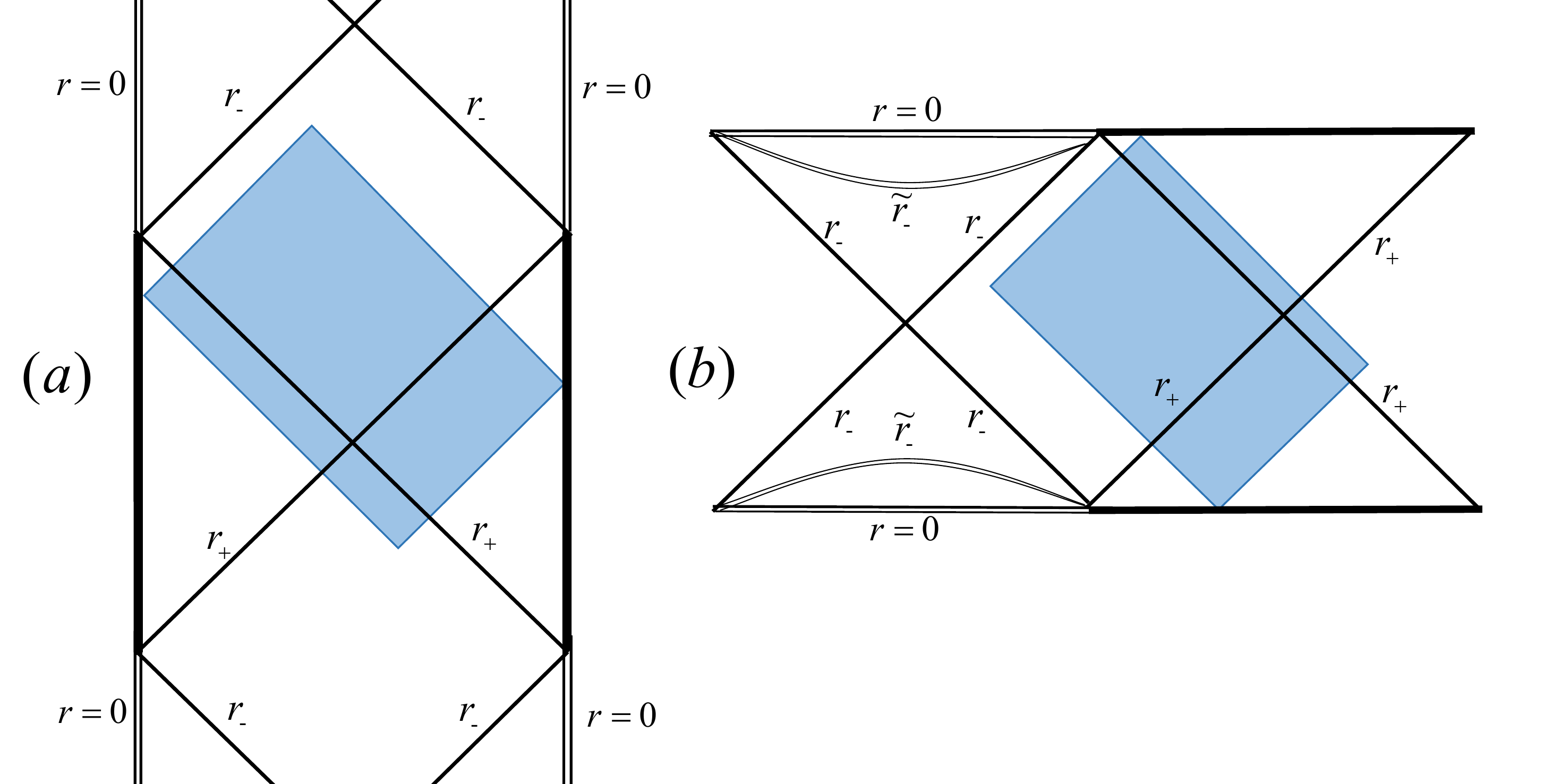}
\caption{(a) The Penrose diagram equipped with the WDW patch for the AdS-RN black hole. The singularity $r=0$ is presented with double line; (b) The Penrose diagram equipped with the WDW patch for the charged Gauss-Bonnet-AdS black hole. The singularities $r=0$ and $r=\tilde{r}_-$ are presented with double lines. }\label{fig1}
\end{figure*}

In the previous works\cite{A4,Pan,2017hwg,Guo:2017rul,WYY}, the CA conjectures for a variety of gravitational theories in the black holes with either the single or multiple horizons have been investigated.
Those results lead to a natural conjecture that the action growth rate for a single horizon black hole can be obtained by taking the limit of its corresponding multiple horizons into one, $e.g.$, the D-dimensional RN-AdS black hole, Rotating/charged BTZ black hole, Kerr-AdS black hole, and Charged Gauss-Bonnet-AdS black hole in \cite{A4}, as well as the rotating BTZ black hole in critical gravity which we will shown in \ref{case2}.  The Multiple-horizon black hole is a more general setting as it reduces to the
single horizon case by the limiting process.
Therefore, in this paper, we only focus on the multiple-horizon black hole case. There are two typical cases: one is the RN-AdS black hole case with the timelike infinity, and the other is the charged Gauss-Bonnet-AdS black hole with the spacelike infinity. The line element is described by
\ba\label{ds2}
d s^2=-f(r)dt^2+\frac{dt^2}{f(r)}+r^2d\W^2_{k,D-2}\,,
\ea
where
\ba\begin{aligned}
f(r)&=\frac{r^2}{L^2}+k-\frac{\w^{D-3}}{r^{D-3}}\,\ \ \ \,\text{and}\\
f(r)&=k\pm\frac{r^2}{2\tilde{\a}}\lf(1-
\sqrt{1+4\tilde{\a}\lf(\frac{m}{r^{D-1}}-\frac{1}{L^2}
-\frac{q^2}{r^{2D-4}}\rt)}\rt)\nn
\end{aligned}\ea
are the blackening factors for the RN-AdS and the charged Gauss-Bonnet-AdS black hole, individually. Their Penrose diagrams with the WDW patch are shown in \fig{fig1}.

Moreover, the results in \cite{A4,Pan,2017hwg,Guo:2017rul,WYY} suggested an expression of the action growth
\ba\label{didt}
\frac{dI}{dt}=\lf(M-\W_{+} J-\m_{+} Q\rt)-\lf(M-\W_{-} J-\m_{-} Q\rt)
\ea
for a multiple-horizon black hole in a guage-gravity theory. Could the form of this equation be held with the presence of the interaction mater field? To answer this question, in the following, we will calculate the action growth rate for any stationary multiple-horizon black hole, and then compare our result with Eq.\eq{didt}.
It is worth mentioning that these quantities in \eq{didt} might be expressed as the conserved charges which are generated by corresponding Killing vectors in the stationary spacetime by the Iyer-Wald formalism. These conserved charges are first derived by Iyer $et\, al.$\cite{IW} in asymptotic flat space and then developed to arbitrary asymptotic one by Wontae Kim $et\, al.$\cite{WSS}. Therefore, the primary goal of this paper is to investigate that whether it is possible to express the late time action growth rate by the Iyer-Wald formalism in a multiple horizons black hole.

The structure of this paper is as follows. In section \ref{Wald}, We briefly review the Iyer-Wald formalism for an invariant theory. In section \ref{agr}, we investigate the action growth rate for a higher curvature gravitational theory coupled with arbitrary matter fields. In section \ref{spc}, we focus on a particular case where the matter fields are composed of a $U(1)$ gauge field and its corresponding complex scalar field. Then, in order to compare with the existing results, we evaluate the action growth rate for the Einstein-Maxwell theory and the rotating BTZ black hole in the critical gravity, respectively. Concluding remarks are given in section \ref{conc}.

\section{Iyer-Wald formalism for an invariant theory}\label{Wald}
In this paper, we would like to use the Noether charge formalism proposed by Iyer and Wald\cite{IW} to investigate the late time action growth rate within WDW patch in a multiple Killing horizons black hole. We know that each symmetry relates to a physical quantity. Moreover, this relation can be constructed by the Iyer-Wald formalism in general invariant theory. Here, we adopt this method to establish a general formalism of the action growth rate in an arbitrary asymptotic spacetime coupling any matter fields.

Firstly, we consider a general diffeomorphism invariant theory derived from a Lagrangian $\bm{L}=\math{L}\bm{\epsilon}$ where the dynamical fields consist of a Lorentz signature metric $g_{ab}$ and other fields $\y$. Following the notation in \cite{IW}, we use boldface letters to denote differential forms and collectively refer to $(g_{ab},\y)$ as $\f$. Generally, the action can be divided into the gravity part and matter part, $i.e.$, $\bm{L}=\bm{L}_\text{grav}+\bm{L}_\text{mt}$. The variation of the gravitational part with respect to $g_{ab}$ is given by
\ba
\d \bm{L}_\text{grav}=\bm{E}_{g}^{ab}(\f)\d g_{ab}+d \bm{\Q}(\f,\d g)\,,
\ea
where $\bm{E}_g^{ab}(\f)$ is locally constructed out of $\f$ and its derivatives and $\bm{\Q}$ is locally constructed out of $\f, \d g_{ab}$ and their derivatives. The equation of motion can be read off as
\ba
\bm{E}_g^{ab}(\f)=\frac{1}{2}T^{ab}\bm{\epsilon}\,,
\ea
where
\ba\label{Tab}
T^{ab}=-\frac{2}{\sqrt{-g}}\frac{\d \sqrt{-g}\math{L}_\text{mt}}{\d g_{ab}}=-g^{ab}\math{L}_\text{mt}-2\frac{\d \math{L}_\text{mt}}{\d g_{ab}}
\ea
 is the stress-energy tensor of the matter fields. Let $\z^a$ be the infinitesimal generator of a diffeomorphism. Exploiting the Bianchi identity $\grad_{a}T^{ab}=0$, one can obtain the identically conserved current for a generic background metric $g_{ab}$ as
\ba\label{J1}\begin{aligned}
\bm{J}[\z]&=\bm{\Q}(\f,\z)-\z\.\bm{L}_\text{grav}
+s_{\z}\.\bm{\epsilon}\,,
\end{aligned}\ea
 where $s^a_{\z}\equiv T^{ab}\z_b$ and  $\bm{\Q}(\f,\z)=\bm{\Q}(\f,\math{L}_{\z}g_{ab})$. Since $\bm{J}$ is closed, there exists a Noether charge $(n-2)$-form $\bm{K}[\z]$ such that $\bm{J}[\z]=d\bm{K}[\z]$. With similar arguments in \cite{IW}, this $(n-2)$-form can always be expressed as
\ba\label{K}
\bm{K}=\bm{W}_c\z^c+\bm{X}^{cd}\grad_{[c}\z_{d]}\,,
\ea
where
\ba\label{Xcd}
\lf(\bm{X}^{cd}\rt)_{c_3\cdots c_n}=-E_R^{abcd}\bm{\epsilon}_{abc_3\cdots c_n}\,,
\ea
 is the Wald entropy density in which $$E_R^{abcd}=\frac{\d \math{L}_\text{grav}}{\d R_{abcd}}\,.$$

Assuming that $\z$ is an asymptotic symmetry, then, there exists a quasilocal ADT conserved charge\cite{WSS} associated with $\z$,
\ba
\d Q_\text{ADT}=2\int_{\inf}\d\bm{Q}_\text{ADT}\,,
\ea
 where $\d \bm{Q}_\text{ADT}=\frac{1}{2}\d \bm{K}[\z]-\z\.\bm{\Q}(\f,\d g)$. Particularly, when $\z^a$ is taken to a rotational Killing vector $\j^a$ in an axisymmetric spacetime, the corresponding conserved charge can be given by
\ba\label{JK}
\math{J}[\j]=-\int_{\inf}\bm{K}[\j]\,,
\ea
which can be interpreted as the angular momentum of the black hole in an arbitrary asymptotic space\cite{WSS}. For a general higher curvature theory, it can be given by
\ba\label{Kj}
\math{J}[\j]=-\int_{\inf}\lf(\bm{X}^{cd}\grad_{[c}\j_{d]}-2\j_{b}\grad_a\bm{X}^{ab}\rt)\,.
\ea
In Einstein gravity, it turns out that
\ba\label{Jj}
\math{J}[\j]=\frac{1}{4\p}\int_{\inf}*d\j\,,
\ea
which is exactly the standard definition of the angular momentum in Einstein theory.

Substitute (\ref{Tab}) into (\ref{J1}), one can obtain
\ba\label{Lz}
\z\.\bm{L}=\bm{\Q}(\f,\z)-d \bm{K}[\z]+\c_\z\.\bm{\epsilon}\,,
\ea
where we denote
\ba
\c^a_\z=-2\frac{\d \math{L}_\text{mt}}{\d g_{ab}}\z_b\,.
\ea
One can verify that $d\lf(\c_{\z}\.\bm{\epsilon}\rt)=0$, so there also exists a $(n-2)$-form $\bm{\L}[\z]$ such that
\ba\label{cz}
\c_{\z}\.\bm{\epsilon}=d\bm{\L}[\z]\,.
\ea
\section{Action growth rate for a higher curvature gravitational theory}\label{agr}
In this section, we consider a stationary black hole with multiple bifurcate Killing horizons, and denote the ``outer" Killing horizon and the first ``inner" Killing horizon to $\math{H}_+$ and $\math{H}_-$, separately. Let
\ba\label{xa}
\x^a_{\pm}=t^a+\W_{\pm}^{(\m)}\j_{(\m)}^a
\ea
be their horizon Killing fields of this black hole which will vanish on the bifurcation surface, where $t^a=\lf(\pd/\pd t\rt)^a$ is the stationary Killing field with unit norm at infinity, $\j_{(\m)}^a$ is the axial Killing fields, and $\W_{\pm}^{(\m)}$ is the angular velocities of the horizon $\math{H}_{\pm}$.

\begin{figure}
\centering
\includegraphics[width=0.35\textwidth]{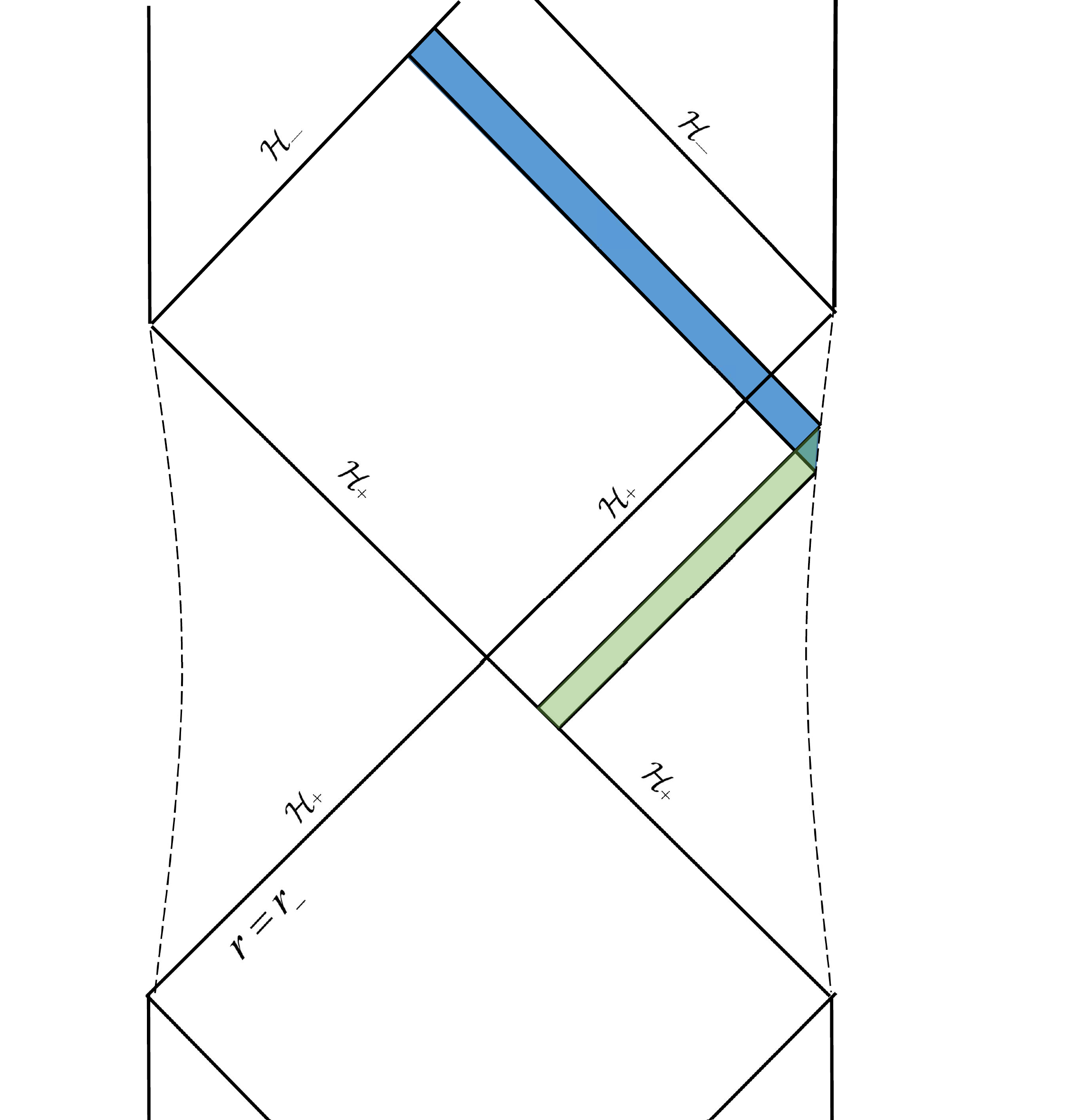}
\caption{Wheeler-DeWitt patch at late time of a multiple Killing horizon black hole, where the dashed lines denote the cut-off surface at asymptotic infinity, satisfying the asymptotic symmetries. }\label{WDW}
\end{figure}

In what follows, we turn attention to the action growth rate within WDW patch at the late time in this stationary black hole for a higher curvature gravitational theory. The corresponding full action can be expressed as\cite{D}
\ba\label{fa}\begin{aligned}
I_\text{full}=&I_\text{bulk}+\int_{\S}\h\bm{X}^{cd}\bm{\epsilon}_{cd}
+\int_{\math{N}} d\l\k \bm{X}^{cd}\bm{\epsilon}_{cd}\\
&+\int_{\math{N}} d\l \bm{\hat{\Y}}\log\lf(l_\text{ct}\Q\rt)\,,
\end{aligned}\ea
where $\l$ is the parameter of the null generator $k^a$ on the null segment, $\bm{\hat{\Y}}=k^a\grad_a\lf(\bm{X}^{cd}\bm{\epsilon}_{cd}\rt)$   and $\Q=\nabla_ak^a$ is the expansion scalar with $l_\text{ct}$ an arbitrary length scale.

We first consider the bulk contribution to the action growth. As illustrated in \fig{WDW}, the bulk contributions only come from the bulk region $\d M_{\pm}$ generated by the Killing vector $\x_{\pm}^a$ through an asymptotic null hypersurface $\math{N}_{\pm}$ which terminates on the portion $\S_{\pm}$ of the Killing horizon $\math{H}_{\pm}$. Then, the action change contributed by the bulk action at the late time approximation can be shown as
\ba\label{Ibulk}
\d I_\text{bulk}=I_{\d M_-}-I_{\d M_+}\,.
\ea
For simplification, we will neglect the index $\{\pm\}$. Turning to the bulk contribution from $\d M$, we have
\ba\begin{aligned}
I_{\d M}&=\int_{\d M}\bm{L}=\d t\int_{\math{N}}\x\.\bm{L}\,,
\end{aligned}\ea
where we use $\d\l=\d t$ by \eq{xa}. The relation $\math{L}_{\x}\f=0$ implies $\bm{\Q}(\f,\x)=\bm{\Q}(\f,\math{L}_\x g)=0$. According to (\ref{Lz}), one can obtain
\ba\label{xL}\begin{aligned}
\int_{\math{N}}\x\.\bm{L}&=-\int_{\math{N}}d \bm{K}[\x]+\int_{\math{N}}d\bm{\L}[\x]\\
&=-\int_{\inf} \bm{K}[\x]+\int_{\S} \bm{K}[\x]+\int_{\math{N}}d\bm{\L}[\x]\\
&=\W^{(\m)}\math{J}_{(\m)}-\int_{\inf} \bm{K}[t]+\int_{\S} \bm{K}[\x]+\int_{\math{N}}d\bm{\L}[\x]\,.
\end{aligned}\ea
Since the Killing horizon contain a bifurcate surface, the first term in \eq{K} vanishes. Then, one can find
\ba
\bm{K}=\bm{X}^{cd}\grad_{[c}\x_{d]}=\k\bm{X}^{cd}\bm{\epsilon}_{cd}
\ea
on the horizon $\math{H}$, where $\bm{\epsilon}_{ab}$ is the binormal of surface $\S$, and $\x^a\grad_a\x^b=\k\x^b$ on the horizon. With this in mind, \eq{xL} becomes
\ba
\int_{\math{N}}\x\.\bm{L}=\W^{(\m)}\math{J}_{(\m)}+\frac{\k}{2\p}S-\int_{\inf} \bm{K}[t]
+\int_{\math{N}}d\bm{\L}[\x]\nn\\
\ea
with $S=2\p\int_{\S}\bm{X}^{cd}\bm{\epsilon_{cd}}$. Considering these relations, (\ref{Ibulk}) becomes
\ba
\frac{dI}{dt}=\left[\W^{(\m)}\math{J}_{(\m)}+\frac{\k}{2\p}S+\L_{\inf}[\x]-\L_{\S}[\x]\right]^{-}_{+}\,,
\ea
where we denote
\ba
\L_{S}[\z]=\int_{S}\bm{\L}[\z]
\ea
with any $(n-2)$-surface $S$. Here the index $\pm$ presents the quantities evaluated at the ``outer" or first ``inner" horizon.

Next, we turn to the boundary and corner contributions to the action growth. Without loss of generality, we shall adopt the affine parameter for the null generator of the null surface; as a consequence, the surface term vanishes on all null boundaries. Meanwhile, we choose $l^a$ as the null generator of the null boundary $\math{N}$, in which $l^a$ satisfies $\math{L}_{\x}l^a=0$. Then, the time derivative of the counterterm contributed by $\math{N}$ vanishes. For the null segment on the horizon $\math{H}$, since $\math{L}_{\x}g_{ab}=0$, we have $\bm{\hat{\Y}}=0$, $i.e.$, this counterterm contribution also vanishes.

The affinely null generator on the horizon can be constructed as $k^a=e^{-\k\l}\x^a=e^{-\k\l}\lf(\frac{\pd}{\pd \l}\rt)^a$. The transformation parameter can be shown as\cite{D}
\ba
\h(\l)=\ln\lf(-\frac{1}{2}k\.l\rt)=-\k\l+\ln\lf(-\frac{1}{2}\x\.l\rt)\,.
\ea
Then, we have
\ba
\frac{d I_\text{corner}}{d t}=\frac{d I_\text{corner}}{d \l}=-\frac{\k}{2\p} S\,.
\ea
 Combining those contributions, we have
\ba\label{dIdt}
\frac{d I}{dt}=\left[\W^{(\m)}\math{J}_{(\m)}+\L_{\inf}[\x]-\L_{\S}[\x]\right]^{-}_{+}\,,
\ea
where $\math{J}_{(\m)}$ denotes the angular momentum associated the Killing vector $\j^a_{(\m)}$ and $\W^{(\m)}_{\pm}$ denotes the velocity of the horizon $\math{H}_{\pm}$.  Moreover, the charge $\L_{\S_{\pm}}$ is the quantity which characterizes the matter fields on the ``outer" and ``inner" horizon. Through the explicit calculation in the $U(1)$ gauge matter fields, one can find that these charges become the quantities associated with the chemical potentials and charges on the corresponding horizon.

This is a general expression of the late time action growth rate within WDW patch on the arbitrary asymptotic spacetime coupled with any matter fields. If this is a pure gravitational solution, this formalism will cover all of the results in the multiple Killing horizons black hole \cite{A4}.

\section{Special case: U(1) gauge matter field}\label{spc}
As an application, we will consider the matter fields which are composed of a $U(1)$ gauge field and its corresponding complex scalar field. And it can be described by the action ansatz
\ba
\math{L}_\text{mt}=\tilde{\math{L}}\lf(F^2\rt)
+\math{G}\left(\y,\left|\math{D}\y\right|^2\right)\,,
\ea
where $F^2=F_{ab}F^{ab}$ with $\bm{F}=d \bm{A}$, and $\math{D}\f=d\y+i \bm{A}\y$ is the covariant derivative of the complex scalar field. The variation of the gauge field part with respect to $A_{a}$ is given by
\ba\label{Lmt}
\d \bm{\tilde{L}}=\bm{\tilde{E}}^a\d A_a+d\bm{\tilde{\Q}}(\f,\d \bm{A})\,.
\ea
The equation of motion can be shown as
\ba\label{EOM2}
\bm{\tilde{E}}^a=j^a\bm{\epsilon}\,,
\ea
where
\ba\begin{aligned}
\bm{\tilde{E}}^a&=-\grad_b\frac{\pd\tilde{\math{L}}}{\pd\grad_b A_a}\bm{\epsilon}\,,
\end{aligned}\ea
and
\ba\begin{aligned}
j^a&=-\frac{\d \math{G}}{\d A_a}=-i\y\frac{\pd \math{G}}{\pd \math{D}_a\y}+i\y^*\frac{\pd \math{G}}{\pd \lf(\math{D}_a\y\rt)^*}
\end{aligned}\ea
is the $n$-current density of the complex scalar field. One can obtain
\ba\begin{aligned}
0&=\bm{\tilde{E}}^a \grad_a \a+d \bm{\tilde{\Q}}(\f,d \a)\,,
\end{aligned}\ea
when considering a gauge transformation $\d \bm{A}=d\a$ with an arbitrary scalar field $\a$ and substituting it into \eq{Lmt}. By using the conserved relation $\grad_{a}j^a=0$ and the equation of motion \eq{EOM2}, we can also define a conserved current of the gauge field
\ba\label{Kj}
d\bm{\tilde{K}}[\a]=\bm{\tilde{J}}[\a]=\a j\.\bm{\epsilon}+\bm{\tilde{\Q}}(\f,d \a)\,.
\ea
According to (\ref{Lmt}), one can further obtain
\ba\label{Kt}
\bm{\tilde{K}}_{c_3\cdots\c_n}[\a]
=\frac{\a}{2}\frac{\pd\tilde{\math{L}}}{\pd \grad_aA_{b}}\bm{\epsilon}_{abc_1\cdots c_n}\,.
\ea
Setting $\a=1$ and denoting $\bm{\tilde{K}}=\bm{\tilde{K}}[1]$, one can define the charge of the gauge field in the spatial region which is bounded by $(n-2)$-surface $S$ as
\ba\label{Qcharge}
Q_S=\int_{S}\bm{\tilde{K}}\,.
\ea
Then, we turn to the late time growth rate of the action. For the last term in \eq{dIdt}, we have
\ba\begin{aligned}
\c_{\x}^a&=-2\frac{\pd \math{{L}}_\text{mt}}{\pd g_{ab}}\x_b\\
=&-\frac{\pd \math{\tilde{L}}}{\pd \grad_{a}A_{c}}F^b{}_c\x_b+A^b\x_bj^a\\
=&\frac{\pd \math{\tilde{L}}}{\pd \grad_{a}A_{c}}\grad_c\lf(A_b\x^b\rt)+A^b\x_bj^a\\
=&-\grad_c\lf(\frac{\pd \math{\tilde{L}}}{\pd \grad_{c}A_{a}}A_b\x^b\rt)+\grad_c\lf(\frac{\pd \math{\tilde{L}}}{\pd \grad_{c}A_{a}}\rt)A_b\x^b+A^b\x_bj^a\\
=&-\grad_c\lf(\frac{\pd \math{\tilde{L}}}{\pd \grad_{c}A_{a}}A_b\x^b\rt)
\end{aligned}\ea
where we have used the symmetric property, $i.e.$, $\math{L}_{\x}\f=0$, as well as the equation of motion \eq{EOM2}. Considering \eq{cz} and \eq{Kt}, we have
\ba
\bm{\L}[\x]=-\bm{\tilde{K}}[\F]=-\F\bm{\tilde{K}}
\ea
where $\F=-\x^aA_a$ is the gauge field potential. By using the smoothness of the pullback of $A_a$ and the stationary condition, one can show that $\F_{\math{H}}=-\left.\x^aA_a\right|_{\math{H}}$ is constant in the portion of the horizon to the future of the bifurcation surface. If we assume that the black hole is asymptotic static, we have $\left.A_a\j^a\right|_{\inf}=0$. Then, the action growth rate within the WDW patch can be shown as
\ba\label{dIdt1}
\frac{d I}{dt}=\left[\W^{(\m)}\math{J}_{(\m)}+\F_{\math{H}}Q_{\S}\right]^{-}_{+}\,.
\ea
Throwing out the complex scalar field, this expression will go back to the result \eq{didt} in the sourceless gravity-Maxwell case\cite{A4}. According to \eq{Kj}, one can find that
\ba
Q-Q_{\S}=\int_{\math{N}}j\.\bm{\epsilon}\,,
\ea
with $Q=\int_{\inf}\bm{\tilde{K}}$ the total charge of the black hole. This expression illustrates the effect of the complex scalar field. It shows the difference between \eq{didt} and \eq{dIdt1}, where $Q$ in \eq{didt} is the total charge of the black hole, while $Q_{\S}$ in \eq{dIdt1} is only the charge inside the corresponding horizon.

\subsection{Explicit case 1: Einstein-Maxwell theory}
To compare with the existing results, let us first consider a $D$-dimensional Einstein-Maxwell gravity. The bulk action can be written as
\ba
I=\frac{1}{16\p}\int d^Dx\sqrt{-g}\lf(R-2\L-F^2\rt)\,.
\ea
Then, the corresponding quantities in \eq{dIdt1} can be expressed as
\ba\begin{aligned}
\F_{\math{H}}&=-\left.\x^aA_a\right|_{\math{H}}\,,\\
Q_{\S}&=Q=\frac{1}{4\p}\int_{\inf} *\bm{F}\,,
\end{aligned}\ea
which are the standard definitions of the chemical potential at each horizon and the electric charge of the black hole, separately. It means that our final formalism \eq{dIdt1} will give the same result in \cite{A4} for the RN-AdS black hole, rotating/charged BTZ black hole as well as the Kerr-AdS black hole in Einstein-Maxwell theory.
\subsection{Explicit case 2: Rotating BTZ black hole in critical gravity}\label{case2}
Finally, to compare our result with the higher curvature case, we consider a 3-dimensional rotating BTZ black hole in critical gravity whose action is given by
\ba\label{action}
I=\frac{1}{16\p}\int d^3\sqrt{-g}\lf[R-2\L-\frac{1}{m^2}\lf(R^{ab}R_{ab}-\frac{3}{8}R^2\rt)\rt]\,,\nn\\
\ea
And the corresponding line element can be expressed by
\ba\label{ds2}
ds^2=-f(r)dt^2+\frac{dr^2}{f(r)}+r^2\lf(d\f+\frac{4J}{r^2}dt\rt)^2\,,
\ea
with the blackening factor
\ba
f(r)=\frac{r^2}{L^2}-8M+\frac{16J^2}{r^2}\,.
\ea
Then, one can obtain the Ricci tensor,
\ba\label{Rab}
R_{ab}=-\frac{2}{L^2}g_{ab}\,.
\ea
According to the Eqs.\eq{Xcd}, \eq{Rab} and the action \eq{action}, one can further obtain
\ba\label{ERabcd1}
E_R^{abcd}=\frac{1}{16\p}\lf(1-\frac{1}{2L^2m^2}\rt)g^{c[a}g^{b]d}\,.
\ea
The parameters $M$ and $J$ can be expressed by the outer and inner horizon $r_\pm$,
\ba
M=\frac{r_+^2+r_-^2}{8L^2}\,,\ \ J=\frac{r_+r_-}{4L}\,.
\ea
By Eqs.\eq{JK},\eq{Kj} and \eq{ERabcd1}, we can obtain
\ba
\math{J}[\j]=J\lf(1-\frac{1}{2L^2m^2}\rt)\,.
\ea
From the line element \eq{ds2}, the angular velocity can be written as
\ba
\W_{\pm}=\frac{4J}{r^2_{\pm}}\,.
\ea
With these in mind, the action growth rate can be expressed as
\ba\begin{aligned}
\frac{dI}{dt}&=4J^2\lf(\frac{1}{r_-^2}-\frac{1}{r_+^2}\rt)
\lf(1-\frac{1}{2L^2m^2}\rt)\\
&=\frac{r_+^2-r_-^2}{4L^2}\lf(1-\frac{1}{2L^2m^2}\rt)
\end{aligned}\ea
Then, the single horizon result reduces to
\ba
\frac{dI}{dt}=2 M\lf(1-\frac{1}{2L^2m^2}\rt)=2M_\text{ADM}\,,
\ea
by the limiting process $r_-\to0$, in which
\ba
M_\text{ADM}=M\lf(1-\frac{1}{2L^2m^2}\rt)
\ea
is the ADM mass for the critical gravity\cite{OE,2017hwg}. And this is exactly the results derived in \cite{2017hwg}.
\section{Conclusion}\label{conc}
In conclusion, we have obtained a general expression \eq{dIdt} of the action growth rate within the WDW patch in a multiple Killing horizons black hole for a higher curvature theory of gravity. The final expression only relies on the Noether charge of the spacetime on the Killing horizon or asymptotic infinity. This result is universal and can be considered as a generalization of the asymptotic AdS to the arbitrary asymptotic one. Finally, we used this formalism to evaluate the late time action growth rate for the $U(1)$ invariant matter fields. To compare with the existing results, we studied the action growth rate for the Einstein-Maxwell theory and the rotating BTZ black hole in the critical gravity, respectively. Then, we found that it covers all of the results in previous letters\cite{A4,Pan,2017hwg,Guo:2017rul,WYY}.

\section*{acknowledge}
 This research was supported by NSFC Grants No. 11775022 and 11375026.


\begin{thebibliography}{100}
\bibitem{1} L. Susskind, Fortsch. Phys. {\bf64} 24(2016).
\bibitem{2} S.Aaronson, arXiv:1607.05256.
\bibitem{4} L. Susskind, Fortsch. Phys. {\bf64}, 24(2016).
\bibitem{5} D. Stanford and L. Susskind, Phys. Rev. D {\bf90}, 126007(2014).
\bibitem{BrL} A. R. Brown, D. A. Roberts, L. Susskind, B. Swingle and Y. Zhao, Phys. Rev. Lett. {\bf116} 191301(2016).
\bibitem{BrD} A. R. Brown, D. A. Roberts, L. Susskind, B. Swingle and Y. Zhao, Phys. Rev. D {\bf93} 086006(2016).
\bibitem{D1} D. A. Roberts, D. Stanford and L. Susskind, JHEP {\bf1503} (2015)
\bibitem{D2} L. Susskind and Y. Zhao, arXiv:1408.2823.
\bibitem{D3} R. G. Cai, M. Sasaki and S. J. Wang, Phys.\ Rev.\ D {\bf 95}, 124002(2017)
\bibitem{D4} L. Lehner, R. C. Myers, E. Poisson and R. D. Sorkin, Phys. Rev. D {\bf94}, 084046(2016).
\bibitem{D5} R. A. Jefferson and R. C. Myers, JHEP {\bf1710} 107(2017).
\bibitem{D6} K. Hashimoto, N. Iizuka and S. Sugishita, Phys. Rev. D {\bf96}  126001(2017).
\bibitem{D7} D. Carmi, S. Chapman, H. Marrochio, R. C. Myers and S. Sugishita, JHEP {\bf1711} 188(2017).
\bibitem{D8} A. Reynolds and S. F. Ross, Class. Quant. Grav. {\bf34},175013(2017).
\bibitem{D9} S. Chapman, H. Marrochio and R. C. Myers, JHEP {\bf1701}  062(2017).
\bibitem{D10} D. Carmi, R. C. Myers and P. Rath, JHEP {\bf1703} 118(2017).
\bibitem{D11} B. Czech, Phys. Rev. Lett. {\bf120} 031601(2018).
\bibitem{D12} P. Caputa, N. Kundu, M. Miyaji, T. Takayanagi and K. Watanabe, JHEP {\bf 1711} 097(2017)
\bibitem{D13} M. Alishahiha, Phys. Rev. D {\bf92}  126009(2015).
\bibitem{D14} P. Caputa, N. Kundu, M. Miyaji, T. Takayanagi and K. Watanabe, Phys. Rev. Lett. {\bf119}, 071602(2017).
\bibitem{D15} A. R. Brown and L. Susskind, Phys. Rev.  D {\bf 97} 086015(2018).
\bibitem{D16} C. A. Agon, M. Headrick and B. Swingle, arXiv: 1804.01561.
\bibitem{D17} O. Ben-Ami and D. Carmi, JHEP {\bf1611}, 129 (2016).
\bibitem{D18} S. Chapman, M. P. Heller, H. Marrochio and F. Pastawski, Phys. Rev. Lett. {\bf120} 121602(2018).
\bibitem{D19} Y. Zhao,  Phys. Rev. D {\bf 97} 126007(2018).
\bibitem{D20} Z. Fu, A. Maloney, D. Marolf, H. Maxfield and Z. Wang, JHEP {\bf02} 072(2018).
\bibitem{D21} L. Hackl and R. C. Myers,  arXiv:1803.10638.
\bibitem{D22} M. Alishahiha, A. Faraji Astaneh, M. R. Mohammadi Mozaffar and A. Mollabashi, arXiv:1802.06740.
\bibitem{D23} J. Couch, S. Eccles, W. Fischler and M. L. Xiao, JHEP {\bf1803} 108(2018).
\bibitem{Huang:2016fks} H.~Huang, X.~H.~Feng and H.~Lu, Phys.\ Lett.\ B {\bf 769}, 357(2017).
\bibitem{D24} B. Swingle and Y. Wang, arXiv:1712.09826.
\bibitem{D25} M. Moosa, JHEP {\bf1803} 031(2018).
\bibitem{D} J Jiang, H Zhang, arXiv:1806.10312.
\bibitem{A2} D. Carmi, R. C. Myers, and P. Rath, JHEP {\bf03} 118(2017).
\bibitem{A5} Luis Lehner, Robert C. Myers, Eric Poisson, and Rafael D. Sorkin, Phys. Rev. D{\bf94} 084046, 2016.
\bibitem{B1} D. Carmi, S. Chapman, H. Marrochio, R. C. Myers and S. Sugishita, JHEP {\bf1711} 188(2017).
\bibitem{F1} S. Chapman, H. Marrochio and R. C. Myers, JHEP {\bf 1806} 046(2018).
\bibitem{F2} S. Chapman, H. Marrochio and R. C. Myers, arXiv:1805.07262.
\bibitem{D26} B. Chen, W. M. Li, R. Q. Yang, C. Y. Zhang and S. J. Zhang, arXiv:1803.06680.
\bibitem{A4} R. G. Cai, S. M. Ruan, S. J. Wang, R. Q. Yang, and  R. H. Peng,  JHEP {\bf09} 161(2016).
\bibitem{Pan} W. J. Pan and Y. C. Huang, Phys.\ Rev.\ D {\bf 95} 126013(2017).
\bibitem{Guo:2017rul} W.~D.~Guo, S.~W.~Wei, Y.~Y.~Li and Y.~X.~Liu,  Eur.\ Phys.\ J.\ C {\bf 77}, 904(2017).
\bibitem{WYY} P. Wang, H. Yang, and S. Ying, Phys. Rev. D {\bf96} 046007(2017).
\bibitem{2017hwg} M.~Alishahiha, A.~Faraji Astaneh, A.~Naseh and M.~H.~Vahidinia, JHEP {\bf 1705}, 009 (2017).
\bibitem{IW} V. Iyer and R.M. Wald, Phys. Rev. D {\bf50}, 846(1994).
\bibitem{WSS} W. Kim, S. Kulkarni and S. H. Yi, Phys. Rev. Lett.  {\bf 111} 081101(2013) .
\bibitem{OE} O. Hohm, E. Tonni, JHEP {\bf04} 093(2010).
\end{thebibliography}
\end{document}